# Teaching Introductory Electrical Engineering Course to CS Students in a Russian University

V.V.Tregub

Novosibirsk State University

vasilich@tregub.ru

May, 2011

## Abstract

This article is about the author's experience with developing and teaching an introductory electrical engineering course for students of Faculty (department) of Information Technology of a Russian university. The curriculum of this department conforms to typical computer science curricula of US engineering schools with a noticeable omission of comparable electrical engineering courses.

When developing the course, I did my best to pay attention to learning preferences of the department's student body. I also hoped to contribute to a degree to meeting labor market demands for developers of electrical engineering CAD software. As for inspiration, I was enchanted with ideas of the Mead & Conway revolution, albeit indirectly related to my enterprise.

-----

# Об опыте преподавания курса "Введение в электронику" будущим программистам

В.В.Трегуб

Новосибирский Государственный Университет

vasilich@tregub.ru

май 2011 года

В рамках учебного плана факультета информационных технологий российского университета, разработан и выдан студентам вводный курс электронной инженерии. Учебный план этого факультета в отношении курсов программирования и информатики примерно соответствует куррикулуму факультетов компьютерных наук (computer science departments) высших инженерных школ Соединенных Штатов.

В разработанном курсе учтены интересы студентов данного факультета, как эти интересы уяснил для себя автор в ходе пятилетней преподавательской работы, и потенциальные потребности рынка труда в разработчиках программного обеспечения для электронных САПРов. Косвенное влияние на разработанный курс оказало изучение автором документов и свидетельств участников «революции Мида и Конвей» в области проектирования СБИС (http://ai.eecs.umich.edu/people/conway/VLSI/VLSIarchive.html). Влияние названо здесь косвенным по единственной причине – руководство факультета пригласило автора для работы над курсом «аналоговой» электроники (в терминологии русской высшей школы). На факультете уже был курс «цифровой» электроники, предположительно охватывающий вопросы проектирования СБИС. С этой оговоркой, вдохновленность идеями «революции Мида и Конвей» в моей работе была определяющим фактором. Я строил свой курс таким образом, чтобы вписать в дальнейшем его в единый учебный план курсов электронной инженерии на факультете, когда и если руководство факультета решит завести специализацию по электронной инженерии.

### Постановка задачи

Дано: Вы взялись преподавать вводный курс электронной инженерии студентам-первокурсникам, пришедшим в ВУЗ с намерением изучать программирование, программную/системную инженерию, информатику и подобные предметы. В их учебном плане действительно много содержательных предметов этого цикла, причем с первого семестра первого курса.

Задача: выдать им курс Введение в электронную инженерию таким образом, чтобы этот курс

- не был изолирован от остальной части учебного плана,
- был интересен большинству студентов,
- составлял завершенный «модуль» в том смысле, что не оставлял бы «хвостов» в представлениях студента о предмете в рамках изученного материала
- и не отнимал учебного времени сверх определенного учебным планом (один семестр, 68 часов по расписанию плюс рекомендуемые 34 часа самостоятельной работы).

К условию задачи (дано) можно отнести также то, что занятия студентов по этому курсу проводятся в компьютерном классе. На компьютерах установлены средства разработки программ (для языков C++, C#, Java, etc) и САПРы типа OrCAD. Лабораторные работы по схемотехнике и электронике, в их традиционном понимании, учебным планом не предусмотрены и условий для проведения лабораторных или хотя бы демонстраций нет.

В статье я опишу рационализацию постановки задачи, мой способ ее решения и достигнутые результаты. В этом смысле, статью можно считать методической, хотя методика преподавания дисциплины в традиционном понимании не есть предмет этой статьи.

Простота формулировок вышеприведенного списка не исключает различия толкований. Вот мое понимание этих формулировок:

- не был изолирован от остальной части учебного плана –
курс Электронная инженерия должен быть не просто добавлен, но интегрирован в организованную совокупность преподаваемых студенческой группе дисциплин. Эта организованная совокупность составляет академическую основу будущей специальности студентов. Употребляя англоязычный термин (из латыни), куррикулум (curriculum) должен быть цельным, не распадаться на фрагменты. В идеале, дисциплины в составе куррикулума должны давать необходимый и достаточный набор знаний и навыков для работы по выбранной специальности. Стараясь ничего не упустить, надо избавляться от заведомо лишнего учебного материала.

- был интересен большинству студентов –
следствие обязательности курса для всего учебного потока. Имей студенты свободу в построении индивидуального учебного плана, этот пункт выполнялся бы автоматически, либо курса не было бы вообще (на него никто бы не записался).
- составлял завершенный «модуль» –
требование очевидно для учебного плана, в котором этот вводный курс – единственный обязательный курс по электронной инженерии за все время обучения. Я работал с учебным планом именно такого типа.
- последний пункт списка дает представление о количественной сложности этой конкретной попытки дать хоть сколь-либо цельную картину предмета Электронная инженерия. В программах высших инженерных школ предмет Электронная инженерия традиционно включает следующие разделы, часто выделяемые в отдельные курсы: Теория цепей, Электроника/Микроэлектроника (принципы работы устройств – как минимум полупроводниковых диодов и полевых тразисторов), Усилители и генераторы, Обработка сигналов, Схемотехника логических элементов, Проектирование электронной аппаратуры и другие. Учитывая последний пункт самопровозглашенных требований к курсу, пришлось оставить за пределами курса ряд важных для предмета в его целостности тем.

## Рассмотренные и опробованные подходы

Вначале о попытке адаптации учебной программы по электронике и электротехнике, рекомендованной министерством образования в 2000-ом году для специальности 552800 Информатика и вычислительная техника (бакалавры).

Для расшифровки состоящего из назывных предложений текста из раздела ОПД.Ф.02 документа министерства образования (http://www.edu.ru/db/portal/spe/os_zip/552800b_2000.html), я изучил ряд интерпретаций этого текста в учебных программах ВУЗов РФ. Названия курсов Электротехника и Электроника, Теоретические основы электротехники, Основы электроники, Аналоговая электроника и т.п. (источник: сайт министерства образования РФ http://www.edu.ru/db/portal/spe/index.htm и ссылки с этого сайта, сайты ВУЗов, обсуждения с коллегами). Судя по практическим заданиям, опубликованные программы курсов по электронной инженерии в качестве профилирующей дисциплины в основном ориентированы на разработчиков и инженеров по эксплуатации РЭА. Технологии элементной базы и разработки РЭА, представленные в этих курсах, относятся к восьмидесятым годам и более давним временам. В немногочисленных программах курсов по электронной инженерии «для общего развития» (естественно-научные факультеты, мехматы) превалирует теоретический компонент. При сопоставлении с программами для «электронно-инженерных» специальностей, заметно, что курсы по электронной инженерии «для общего развития» составлены как адаптация курсов, преподаваемых на «электронно-инженерных» специальностях. Библиография программ курсов не отличается разнообразием: годы издания рекомендованной литературы если и относятся к нынешнему веку, то это по большей части допечатки литературы «до-компьютерной» эры технологий разработки и изготовления РЭА.

В курсах схемотехники «до-компьютерной» эры уделялось много внимания расчетам параметров и характеристик стандартных схемных решений и скомпонованных из этих решений цепей. Симуляторы электронных цепей автоматизируют эту задачу. Означает ли это, что в курсе, в котором на практических занятиях используют симулятор, можно перенести акцент с «ручного» расчета на компьютерный DC/AC/Transient-анализ? Отвечая на этот вопрос, надо принять во внимание, что в «до-компьютерную» эру к расчетам усилительных каскадов, генераторов, фильтров обращались не только для целей практической разработки и анализа аппаратуры. Плодотворное проектирование требует знания предназначений и принципов работы стандартных схемных решений. В унаследованных курсах «до-компьютерной» эры, это знание усваивается студентами по большей части через изучение методов «ручного» расчета цепей.

Практические занятия вводного курса схемотехники начинаются с решения линейных цепей с резисторами и постоянными источниками тока/напряжения. При «концентрированном» прохождении этой темы (за одно-два занятия), естественно сосредоточить внимание студентов на топологии и назначении цепей. Благодаря появившейся возможности рассчитывать цепь в симуляторе, изучение технических приемов расчета (эквивалентные замены, Нортон/Тевенен, «суперузел» и т.д.) можно отложить на время или полагаться на самостоятельное освоение студентами этих приемов по мере возникновения необходимости.

Использование программного инструмента (симулятора) в заданиях «докомпьютерного» репертуара "как есть", без дополнительной методической проработки, сводит эти задания к демонстрации. Для любой топологии цепи расчет токов/потенциалов, мощности, вольтамперных характеристик, параметрических зависимостей запускается единообразными действиями в пользовательском интерфейсе симулятора. При запуске, симулятор выполняет над введенной студентом схемой нечто вроде Design Rule Checking: расчет не запускается, если какая-либо изолированная ветвь схемы не подключена к земле, есть источники тока на холостом ходу или короткозамкнутые источники напряжения. Это, пожалуй, единственный момент «отладки» проекта для такого рода заданий.

Для будущих разработчиков аппаратуры и чипов, изучение схемотехнических решений может и должно начинаться на самом первом занятии – пусть даже изучаемые на этом занятии цепи содержат лишь резисторы и постоянные источники. Так, параметрический DC-анализ позволяет построить калибровочную кривую для измерений с фиксированным мостиком Уитстона. Различные вариации схемы мостика позволяют компенсировать сопротивление соединительных проводов, измерять малые сопротивления, и т.д. Материала достаточно на занятие и останется на домашнее задание.

Для разработчиков программного обеспечения связь расчета мостиков с их будущей профессией не столь очевидна. Первое занятие, проведенное в том же ключе, что и для «железячников», не прояснит студентам пользу изучения электронной инженерии для специализации в областях информатики и программирования.

Сказанное относится ко всему курсу, не только к первому занятию. Обратимся к теме Переходные процессы. Хоть изучение искажения импульса при распространении по RC-линии и имеет непосредственное отношение к «цифровой» электронике, но чтобы усмотреть здесь связь, надо иметь представление, например, о трассировке проводников в интегральных схемах или о способах борьбы с искажениями на шине. Любое задание на исследование «в лоб», с помощью симулятора, поведения катушки как источника тока и конденсатора как источника напряжения оказывается скорее демонстрацией, чем задачей или хотя бы упражнением, и соответственно мало чему учит.

Есть общее объяснение недостаткам изучения схемотехники и электроники по унаследованной методике и при этом с симулятором, но без «натурных» лабораторных. Вспомним, что будущие инженеры и исследователи, в дополнение к используемым в области электронной инженерии концепциям и подходам «фундаментальных» наук, должны освоить методы собственно электронной инженерии, не менее изощренные и обстоятельные. Объединяют же эти концепции и методы в самостоятельную дисциплину их приложения. Без осведомленности о практическом применении компонентов и цепей, выполнение упражнений на симуляторе не приближает студентов к «электронно-инженерному» пониманию процессов в полупроводниковых приборах и не развивает интуицию и навыки конструирования в области схемотехники столь интенсивно, как следовало бы с учетом напряженного таймлайна курса. С другой стороны, изучение приложений электроники требует знания основ предмета. В «докомпьютерную» эпоху этот круг размыкался «переслоением» теоретического и прикладного материала при присущем той эпохе темпе изучения предмета. Приложения изучались параллельно с основами, что обеспечивало целостное восприятие предмета.

Оценивая по сформулированным в начале статьи рекомендациям для постановки курса: попытка «осовременивания» докомпьютерной методики только за счет введения в курс моделирования на симуляторе не была успешной ни по единому из пунктов. Хуже всего дело обстояло с первым пунктом, «пере-использованием» знаний и навыков из области аналоговой схемотехники в других курсах учебного плана факультета. Попросту говоря, ни в одном из дальнейших курсов не требовалось умения рассчитать или спроектировать электронную цепь, с симулятором или без оного, или хотя бы знать таксономию электронных цепей.

В курсах программирования, математики и физики на первом году обучения вводят большое количество абстрактных понятий, необходимых при изучении этих дисциплин по любой методике. Не то чтобы электронная инженерия возможна без абстрагирования, но, по осмыслении опыта использования симулятора в классе, у меня появилась идея «снижения уровня дополнительного абстрагирования». Того абстрагирования, которое самим фактом применения симулятора в учебном процессе добавляется в еще один предмет обязательного набора дисциплин. Идея вызревала, и я решился на ее внедрение, когда понял, как это можно сделать без возвращения к «докомпьютерной» методике.

Но вначале о параллельно предпринятой попытке ввести в курс элементы подхода Агарваля и Лэнга (Profs. Anant Agarwal and Jeffrey H. Lang). Профессоры Агарваль и Лэнг ведут в Массачусетском Институте Технологии на факультете EECS курс 6.002 Цепи и Электроника (http://ocw.mit.edu/courses/electrical-engineering-and-computer-science/6-002-circuits-and-electronics-spring-2007). В предисловии к своему учебнику по данному курсу (http://elsevier.insidethecover.com/searchbook.jsp?isbn=9781558607354), авторы упоминают о двух целях постановки курса: объединить в одном курсе изучение схемотехники и электроники (под электроникой Агарваль и Лэнг подразумевают здесь принципы работы полупроводниковых приборов) и четко связать содержание курса с современными решениями для цифровых и аналоговых систем. На первый взгляд, эти цели отличаются от задач постановки курса, сформулированных в начале моей статьи. Но ознакомьтесь с «расшифровкой» этих целей. Так, аргументируя необходимость объединения курсов схемотехники и электроники, авторы указывают на необходимость «интенсификации» курса (последний пункт моего списка).

Курс 6.002 преподается второкурсникам (sophomores), и среди прочего должен, по утверждению авторов, обеспечить студентов логическим обоснованием «перехода от мира физики к миру электроники и вычислений». В подтверждение этой претензии, учебник содержит в приложении рассуждения авторов о том, как соотносятся приближение сосредоточенных параметров и уравнения Максвелла. Но в материалах собственно курса 6.002, «университетская» физика присутствует разве что в виде формулы Шокли для вольтамперной характеристики диода. Упоминание о переходе от физики к электронике, решил я, лишь отвечает – без ощутимых последствий для содержания курса – представлениям авторов о существовании естественнонаучных основ электроники, и поэтому все же попробовал воспользоваться практически значимой составляющей курса 6.002. В частности, мне показалось уместным представить будущим компьютерным специалистам искусство схемотехники через реализацию элемента логики или ячейки динамической памяти – в дополнение к усилителям, фильтрам и генераторам. Я понимал также, что заимствовать только изолированные примеры вне контекста курса не имеет смысла. В отличие от неубедительной для меня необходимости демонстрировать во вводном курсе «переход от мира физики к миру электроники и вычислений», педагогическая необходимость концепции «цифровой абстракции» сомнения не вызывала, и я ввел в курс также изучение «цифровой абстракции».

Реакция студентов на инновации оказалась более чем сдержанной. К примеру: изучаем поведение цепей в режиме большого сигнала: расчет по модели транзисторов, «измерение» с помощью симулятора, анализ. Традиционно для исследования я предлагал выбрать однотранзисторный каскад, «длиннохвостую» пару, ОУ с нелинейной ООС etc. Под влиянием курса 6.002 добавил в список логический элемент – инвертор КМОП. В начале семинара убедился, что студенты понимают схемотехнику инвертора (изучали на предыдущих занятиях) и значение инвертора для «цифровой» электроники. Напомнил о связи передаточной характеристики с выбором пороговых значений логических напряжений. Несмотря на явную «подсказку» со стороны преподавателя (меня), для анализа и построения графиков передаточной характеристики цепей большинство студентов выбрали схему включения транзистора с общим истоком, шедшую первой в предложенном списке. Никто из группы не взялся детально изучить инвертор: передаточную характеристику, зависимость от отношения ширин каналов, etc.

Ни аналоговые усилители, ни внутренняя схемотехника логических элементов не являются непосредственными объектами исследования или оперирования ни для программистов, ни для «информатиков», ни в теории, ни на практике. Если необходимость изучения этих вещей представляется студентам в равной степени абстрактной, то ссылка на то, что «в дальнейшей работе вам это пригодится» не просто не сработает, но будет «дачей заведомо ложных показаний». В учебном процессе не отпущено время, да и не ставится цели, чтобы снабдить эту абстракцию достаточной силы выразительностью. Схема инвертора для студента компьютерного факультета ничуть не «роднее», чем схема диффкаскада, и тот факт, что инверторов в чипах и цепях компьютера куда как больше, чем диффкаскадов, не добавляет (и не может добавить) студенту мотивации при изучении схемотехники инверторов. В этом свете и усилители, и триггеры суть абстрактные понятия как до курса, так и после – если изучать их per se, вне связи с системами (измерительными или вычислительными, например).

Такого рода абстракции отдаляют новичка от изучаемого материала ничуть не в меньшей степени, чем изучение цепей в «моделирующем реальность» симуляторе вместо «реалистичного» брэдборда. Пояснение для облегчения беглого просмотра этой статьи: «уровни абстрагирования» курса 6.002 не имеют непосредственного отношения к обсуждаемой здесь педагогической проблеме абстрагирования. Скорее, авторы курса 6.002 экстраполировали по своему разумению иерархию уровней абстрагирования из методологии проектирования СБИС (вентильный/RTL/поведенческий уровни описания систем). Во вводном курсе, методика преподавания схемотехники встречается с абстрагированием совсем другого рода.

Методический опыт с подходом курса 6.002 укрепил меня в мысли о необходимости «снижения уровня дополнительного абстрагирования».

### Анализ ситуации

«Необходимость снижения уровня дополнительного абстрагирования» была подсказана все же в большей степени интуицией, чем опытом, ведь идея родилась чуть ли не в первый год моей «педагогической практики». Тогда же или чуть позже я понял, что стоило бы выяснить совсем просто формулируемые вопросы: чему мы учим фактически? Что хотят знать и уметь студенты или что захотят узнать и чему научиться, если удастся увлечь их предметом во вводном курсе? Что рекомендовать им узнать и чему научиться, учитывая нынешнюю ситуацию в отрасли и их специализацию? И что из необъятной тематики предмета включать во вводный курс?

Это были совсем не те вопросы, которые полагалось задавать в рамках концепции «удовлетворения социального заказа». Похоже, и методкомиссию не интересовала методика преподавания вводного курса электронной инженерии. В классе, однако, пройти мимо этих вопросов было невозможно – разве что подождать, когда «социальный заказ» бизнеса и государственных структур ВУЗам на удовлетворение потребностей «новой экономики» окажет влияние на учебные планы инженерных факультетов и вопросы решатся кем-то или чем-то со стороны, а то и просто отпадут по ходу дела.

Я начал отвечать на первый вопрос (чему мы учим фактически?) «от печки» – анализируя материал первого занятия по схемотехнике: закон Ома, правила Кирхгофа, резистор, источники постоянного тока/напряжения. Я пишу «резистор» вместо «сопротивление», подчеркивая, что вначале обычно представляют материальный объект: дискретный компонент или проводящую область на чипе. В электрических цепях используется свойство резистора «сопротивляться» току. Главный параметр резистора – омическое сопротивление. По мере изучения необходимого материала, далее в курсе рассматриваются и «неидеальности» резистора – ограничения по рассеиваемой мощности, нелинейность, паразитные емкость/индуктивность и т.д.

После резисторов и источников изучают в той или иной последовательности реактивные, нелинейные, активные компоненты, осцилляторы, линии, волноводы. Параллельно с изучением принципов работы устройств изучается схемотехника. Физическое представление о «принципе работы» резистора также развивается, в частности, в продвинутых курсах изучали модель Друдэ. К тому моменту в курсе, когда «добирались» до дифференциального сопротивления эмиттерного перехода или до импеданса, резистор и сопротивление были уже достаточно привычны – и объект, и понятие.

При изучении электроники с «натурными» лабораторными упомянутый «уровень дополнительного абстрагирования» если и возникал, то не создавал особых проблем. Процесс абстрагирования шел плавно и стартовал от «реальных» вещей. Компонент «резистор» на лабораторных материален. Развитие представления об этой вещи идет параллельно вдоль «теоретического» и вдоль «практического» направлений. Студент узнает «схемотехнические свойства» резисторов в линейных цепях, решая задачи на эквивалентные цепи, принцип суперпозиции и преобразования источников. В практикуме при отладке усилителя с температурной стабилизацией студент наблюдает физическое свойство температурной зависимости сопротивлений. Курс может включать лабораторную по изучению вклада теплового шума сопротивлений в усиленный сигнал.

Затем, при изучении цепей переменного тока, вводится импеданс как обобщение понятия сопротивления. Импеданс – абстракция, но эта абстракция в одном шаге от «почти реального» сопротивления пассивного элемента, резистора. Другие «одношаговые» абстракции – отрицательное дифференциальное сопротивление (на участке ВАХ диода Ганна или туннельного) или схемотехнически реализованное отрицательное сопротивление порта. Сильная сторона «поэтапного» развития понятия «сопротивление» в курсе унаследованной методики состоит в историчности и естественности, с точки зрения преподавателей.

В интегральной схемотехнике – предположительно наиболее интересной теме для студентов компьютерного факультета – резисторы «не явлены студенту в непосредственном чувственном восприятии». Более того, при продвижении в субмикронную область доля резисторов как специально создаваемых структур на чипе уменьшается. Стоит ли извлекать из анналов электроники резистор для введения понятия омического сопротивления?

Совсем обойтись без материального референта понятия «сопротивление» не лучшее решение. Но в этом и нет нужды. Из школы, житейского опыта и через телевизор/Интернет любознательные так или иначе получают некоторое представление об электрическом сопротивлении. В университетском курсе это представление развивают, чтобы студент мог умело обращаться с сопротивлением: например, как с параметром вольтамперной характеристики ребра графа схемы цепи. Производство дискретных резисторов или формирование резистивных структур на чипе – все это может быть темой специальных курсов, однако во вводном курсе электронной инженерии необходимо начать со схемотехнических аспектов понятия сопротивления, включая цели применения резисторов в цепях.

Для чего педагогическая практика не придумала пока лучшего способа, чем расчет и анализ различных топологий: делители тока/напряжения, мостики, лестничная цепь, звезда/треугольник, а также задачи по темам преобразование источников, принцип суперпозиции, несть им числа.

Что делает разумное и свободное существо, принуждаемое к выполнению множества единообразных операций? Либо уклоняется всеми способами от этой рутины, либо пользуется инструментом. Оба варианта поведения наблюдают преподаватели вводного курса электронной инженерии. «Инструментом» может быть списывание, поиск решений в Интернет, использование готовых утилит (симуляторов) для решения задач. Конкретное поведение – уклонение, выбор инструмента – зависит от обстоятельств, в частности, от способа контроля учебного процесса.

Что сделает разумное существо, <u>осознавшее</u> необходимость выполнения множества единообразных операций, если только это не фитнесс и не развитие мелкой моторики движений? Я выдвигаю следующий Ansatz: существо изготовит для этого орудие труда, tool. Или даже изобретет. По крайней мере, попытается.

Но именно этого, по большому счету, мы и добиваемся от студентов. Как же на практике сподвигнуть студентов компьютерного факультета к изготовлению/изобретению программных инструментов во вводном курсе электронной инженерии? Я понял, что задания с элементами программирования по схемотехнике могут стать отнюдь не популистским средством потрафить студентам в их «программистских» наклонностях, но «подведением» их к осознанному созданию инструмента для индивидуального применения в учебных целях.

И еще: если бы удалось заинтересовать будущих программистов и информатиков изучением электронной инженерии на собственной территории компьютерной науки, это дало бы практически значимое расширение тезиса, приписываемого Дейкстре: Computer science is no more about computers than astronomy is about telescopes. Я не нашел оригинального произведения Дейкстры с цитатой про компьютеры и телескопы, чтобы понять контекст, но сие высказывание, похоже, имеет силу производящей функции высказываний на ту же тему. В частности, я сконструировал (не противоречащее исходному) высказывание computers are no less important for applied computer science than telescopes are for astronomy.

Первое задание на разработку кода, имеющего отношение к расчету электрической цепи, я выдал еще в тот год, когда преподавал в рамках «до-компьютерной» методики. Время от времени я пытался привлечь прилежных студентов, быстро осваивающихся с пользовательским интерфейсом симулятора и не испытывающих затруднений при выполнении символических вычислений в тетради и на доске (и откровенно скучавших на занятиях), к изучению дополнительных тем типа активные фильтры высших порядков, частотные модуляторы/демодуляторы, осцилляторы и ФАПЧ, etc. Аргументированное сопротивление дополнительной учебной нагрузке выражалось типичной фразой: «я пришел на факультет изучать программирование, не аппаратуру».

Действуя прямолинейно, я решил дать задание на расчет цепи, включающее элементы программирования. Есть задача «докомпьютерного» репертуара: вычислить сопротивление между двумя узлами прямоугольной решетки с квадратными ячейками при заданном сопротивлении звена. При использовании симулятора и для небольшой решетки, задача решается «в лоб». Искомое сопротивление определяется «подключением» тестового источника тока к требуемым узлам и «измерением» падения напряжения между этими узлами. Для решетки с большим количеством узлов «ручной» ввод схемы цепи утомителен, и я предложил студенту написать «генератор» нетлиста решеточной цепи резисторов. За два занятия (я не поинтересовался затраченным временем вне класса) студент-первокурсник написал на C и отладил код двух функций, каждая с двумя параметрами – количество узлов решетки по вертикали и количество узлов по горизонтали. Функции выдавали файл с текстом нетлиста. Одна функция подключала периферийные сопротивления к общему узлу, в другой подключений к общему узлу не было. Далее студент руками дописывал в файл нетлиста источник тока, подключенный к нужным узлам, SPICE-инструкции распечатки результатов и анализа по постоянному току, а также заземлял нужные узлы, где необходимо, и «скармливал» файл симулятору.

В качестве следующего шага, я предложил решить в коде всю задачу целиком и обойтись без симулятора. Для задачи вычисления потенциалов в узлах решетки с квадратными ячейками получился, естественно, код решения уравнения Пуассона в конечных разностях. Два типа закодированных студентом функций соответствовали использованию граничных условий Дирихле и Неймана. Помогло ли это упражнение в изучении теории цепей? Несмотря на на первый взгляд отвлекающий экскурс в «методы вычислений», студент в дальнейшем непоколебимо правильно применял правила Кирхгофа и закон Ома для анализа различных цепей, не ошибаясь в выборе знаков, направлений и независимых переменных. Обстоятельства указывали на то, что проделанное упражнение имело к этому отношение.

### Решение и новые проблемы

Следующий год, ведя курс все еще по традиционной методике (с элементами курса 6.002 МИТ), я изучал возможность ввести элементы программирования в задания по теории цепей и электронике на регулярной основе. Выяснилось, что, хотя большинство студентов и имеет опыт программирования, причем не только на «пропедевтических» языках школьной информатики, но и на С, однако написать код для парсинга нетлиста или вывода графика в состоянии лишь единицы (следовательно, включать это в задания – нарушение второго пункта моего самопровозглашенного списка задач по постановке курса), и, скорее всего, это отнимет у них все время, выделенное на курс электронной инженерии (нарушение последнего пункта).

Для изучения основ электронной инженерии, кодирование парсинга нетлиста во вводном курсе не есть задача sine qua non, но каким образом студент мог бы передавать данные о топологии цепи и номиналах функциям своей программы? Я построил следующий сценарий «программистско-ориентированного» изучения схемотехники.

- Преподаватель предоставляет код функции парсинга нетлиста, предназначенной для перевода данных о цепи <u>в удобный для использования при программных вычислениях вид.</u>
- Студент разрабатывает алгоритм расчета цепи с резисторами и источниками, применяя правила Кирхгофа и закон Ома, и далее реализует алгоритм в коде.
- Затем студент изучает поведение реактивных элементов и разрабатывает алгоритмы для расчета переходных процессов и реализует алгоритмы в коде. Программный модуль студента включает служебные функции типа вывод графики, предоставляемые при необходимости преподавателем.
- После (или перед реактивными элементами) идут модели диодов, транзисторов и алгоритмы линеаризации
- И в завершение, насколько позволит время – интеграция алгоритмов в один программный блок, чтобы получить программу расчета цепей, содержащих одновременно и реактивные, и нелинейные элементы; анализ сходимости, точности; тестирование программы расчетом усилительных каскадов и генераторов...

Полученные исподволь в ходе выполнения этих заданий представления о демпфированных колебаниях и резонансах в контурах, о смещении рабочей точки и о слабосигнальном приближении, полагал я, станут «бонусом» с точки зрения студента, и эффективно достигнутой целью курса, с точки зрения преподавателя.

«Удобный для использования при вычислениях вид» имеет информация, представленная в форме массивов структурированных данных (arrays of structs или arrays of objects). Для целей, реально достижимых во вводном курсе, нет необходимости разрабатывать специальные классы. В достаточно полной библиотеке классов есть все функции, которые могут потребоваться студентам в разрабатываемом проекте для работы с данными. Но будет ли по силам первокурсникам работать в парадигме ОО? Не будучи абсолютно уверенным в «законности» замышляемого, я проконсультировался с преподавателем ОО дизайна. Он заверил меня, что все окей, объектное представление данных есть архетип, и может использоваться a priori. Благодаря этому свойству ОО-представления, сильно сбить студентов с толку мне не удастся, тем более, что собственно ОО-конструкции в предполагаемых заданиях используются как данность, и если что, в дальнейших курсах мои упущения легко исправимы, так что «Дерзай!».

Для первого занятия я подготовил заготовку с декларациями структур «сопротивление» и «источник тока», с декларациями массивов компонентов и с функциями парсинга нетлиста и вывода матрицы проводимостей в виде HTML-таблицы. «Схемотехническая» часть задания состояла в предложении «вручную» составить матрицы проводимостей для делителей тока и напряжения, проанализировать процесс их составления, и вывести алгоритм заполнения матрицы и столбца правых частей уравнений для цепи общего вида с сопротивлениями и источниками тока. «Программистская» часть задания – реализовать выведенный алгоритм в коде, чтобы вычислять матрицу проводимостей для любого нетлиста с сопротивлениями и источниками тока [ http://ccfit.nsu.ru/~tregub/eecs/notes1.htm ].

Многие студенты справились с первым заданием еще в классе. В конце занятия я показал, как скопировать выведенную таблицу проводимостей и столбца правых частей из браузера в Excel и с помощью табличных функций Excel решить систему и вычислить значения потенциалов в узлах – с оговорками насчет практической ценности такого метода решения. Студенты уже встречались по жизни с матричными операциями в электронных таблицах. В целом, группа как должное приняла домашнее задание реализовать в коде решение системы линейных алгебраических уравнений (СЛАУ) методом исключения Гаусса, чтобы обходиться в дальнейшем без программы электронных таблиц. Метод Гаусса к этому времени уже «проходили» в курсе линейной алгебры.

Для «обкатки» самодельного симулятора, предложил студентам исследовать с его помощью несколько цепей («лесенка» сопротивлений, квадратная решетка, мостик). На случай, если «самодельный» симулятор не получается или вообще нет желания им заниматься, рекомендовал анализировать предложенные цепи PSPICE-ом или GNUCap-ом, в зависимости от операционной системы на компьютере. В официально утвержденной программе не было темы «разработка симулятора», и я не имел формального права настаивать на выполнении заданий по этой теме.

Начало следующего занятия не указывало на особый успех затеи с «обертыванием» электронной инженерии в программистские задания. Работавшие над кодом студенты «застряли» на реализации функции решения системы линейных уравнений. «Традиционалисты», как и в прошлые годы, боролись с многоэтажными дробями, возникающими при применении метода узловых потенциалов/контурных токов, или с накладками первых опытов ввода схем в симулятор. Вместе с тем, «неприкаянных» в классе стало значительно меньше, чем в предыдущие годы. Конечно, на втором занятии еще нельзя было утверждать, помогла ли новая методика или действуют факторы, не связанные с моей методикой. Я показал статью в википедии про метод исключения Гаусса. Желающим предоставил свою реализацию функции решения СЛАУ. После обсуждения возможных практических применений цепей из домашнего задания, перешли к модифицированному методу узловых потенциалов. Целенаправленно, хоть и без особого энтузиазма, кто-то вернулся к недорешенному домашнему заданию, другие приступили к выполнению нового задания.

Уже на втором занятии в учебных группах стали различимы подгруппа принявших, и подгруппа проигнорировавших мою «инновационную» методику. Даже с учетом малообнадеживающего опыта с первым домашним заданием, результаты работы студентов по «вычислительской» методике на втором занятии позволяли сохранять веру если не в продуктивность, то в небезразличность этой методики для студентов компьютерного факультета.

Количество студентов в «подгруппе энтузиастов» нового метода изучения схемотехники на последующих занятиях не уменьшалось, а я приобретал опыт ведения как бы двух курсов параллельно, одновременно пропагандируя среди коллег точку зрения насчет равных прав на существование двух «путей познания» электронной инженерии, «вычислительского» и «коннекционистского»[1]. Два этих термина всплыли из подсознания при первом разговоре с коллегой на эту тему. Хотелось дать краткое название методике изучения схемотехники через разработку кода симулятора, в отличие от традиционного курса с лабораторными и с символьными расчетами на семинарах. Коллега тогда подсказал мне: для признания методики в профессиональной и академической среде весьма убедительно было бы показать в учебном процессе, что при разработке кода солвера («вычислительская» методика) студенты усваивают тот же набор базовых понятий схемотехники, что и при выполнении лабораторных работ на брэдбордах, хотя и в несколько ином порядке. Традиционную методику я назвал мысленно «коннекционистской», имея в виду, что подготовка к исследованию цепи и на бредборде, и в симуляторе начинается с установления соединений (wire connections) между компонентами цепи.

Когда в курсе началось изучение усилительных каскадов, генераторов и фильтров, я заметил еще один критерий разделения учебных групп в моем курсе на «подгруппы». Первый критерий, напомню, определялся отношением к «вычислительской» методике. Поначалу я видел ситуацию так: мотивированные или амбициозные студенты занимались или по крайней мере предприняли попытку заниматься по «вычислительской» методике. Наблюдение согласовывалось с моим опытом прошлых лет (мантра «я пришел на факультет изучать программирование, не аппаратуру»). Для студентов с невысокой мотивацией задания из унаследованного курса и без «натурных» лабораторных выглядели предпочтительнее, учитывая возможность списать из многочисленных пособий и

---

[1] Слова «вычислительский» и «коннекционистский» я счел возможным употребить, не имея в виду термины теории познания, но и не исключая коннотации. «Коннекционизм» иногда употребляется в значении «коннективизм» (одна из теорий обучения). Интересно было бы сопоставить в рамках педагогической науки две методики изучения электроники, о которых идет здесь речь, но пока это вне моей компетенции, поскольку я прочитал про эти теории только что, при написании этой сноски.

Интернет-ресурсов. Строго говоря, первый критерий не имел отношения к дилемме '«вычислительский» / «коннекционистский» «пути познания» в электронной инженерии', а всего лишь отражал степень заинтересованности студента в курсе.

К середине семестра я обратил внимание на студента, которого задачи из репертуара «вычислительской» методики не интересовали, но который аккуратно выполнял одно за другим ВСЕ мои задания по схемотехнике и электронике, используя готовые инструменты – аналитические вычисления и коммерческий симулятор. Я стал задавать ему на каждом семинаре вопросы «на понимание» и дополнительные задачи, и убеждался, что его владение пройденным материалом превосходит успехи всех одноклассников, включая и тех, кто наиболее продвинулся в разработке симулятора. Увлечение электроникой, занятия в радиокружках студент отрицал. В своем качестве, он не принадлежал ни к одной из двух подгрупп, на которые я мысленно разделил студентов в своем курсе. Пришлось выделить его в подгруппу, состоящую из одного студента.

Я определил способ, каким этот студент изучал предмет, как «коннекционистский», а традиционную методику переименовал в «унаследованную». Для своего внутреннего использования я по крайней мере упорядочил терминологию. Дальнейшая практика преподавания вводного курса электронной инженерии с выбором студентом «пути познания» поможет разобраться, был ли способ изучения электронной инженерии студентом-«коннекционистом» необычным и исключительным для мотивированного студента компьютерного факультета, или, при надлежащей статистике, определенная доля студентов компьютерного факультета предпочтет в курсе электронной инженерии использовать готовые программные инструменты, реализуя творческие импульсы через проектирование инструментов и приборов на аппаратном уровне.

Мышление по аналогии подсказывает, что «вкусы» этой подгруппы студентов в отношении языков программирования могут лежать скорее в области HDL, чем C или Java, но интуиция намекает, что все будет гораздо сложнее. Возможно, методика будет предусматривать разработку студентом по каждой теме курса отдельного DSL (Domain Specific Language) для кодирования задач данной темы, решаемых далее с использованием утилиты, написанной на C++ тем же студентом или его напарником по учебному проекту.

## Заключение

Хотя первый «прогон» курса по «вычислительской» методике не привел к резкому улучшению знаний студентов (по крайней мере, экзамен этого не показал), увеличилось количество студентов, реально работавших и на семинарах, и самостоятельно. Лучший способ оценки качества вводного курса – отслеживание успехов студентов в дальнейших курсах, связанных с электронной инженерией. Для этого предстоит наладить сотрудничество преподавателей курсов электронной инженерии в рамках согласованного подхода к развитию инженерного образования.

Это сотрудничество хотелось бы укрепить взаимодействием с преподавателями математики, физики и численных методов. Естественно сформулировать пожелание, чтобы студенты по окончании курса линейной алгебры были в состоянии закодировать решение СЛАУ, переопределенной системы, найти ту или иную факторизацию положительно определенной матрицы, etc. И ведь вышеупомянутое обоснование «перехода от мира физики к миру электроники», если понимать его прагматически, может содержать изучение квазистационарного приближения электромагнитной теории, необходимого для расчетных задач электроники и развития схемотехнической интуиции. Только учить квазистационарному приближению надо в курсе физики, коль скоро таковой есть в учебном плане факультета.

Возвращаясь к вводному курсу электронной инженерии: в результате своего рода положительной обратной связи, работа над новой методикой повлияла на самого преподавателя (меня) и отразилась на качестве и выдаваемых мною заданий, и моей интерактивной работы со студентами. Сравнивая свои учебные материалы по новой и «унаследованной» методике, я вижу улучшение качества в вещах, не связанных напрямую с заданиями по разработке симулятора, методами вычислений и программированием. Новый курс стал, прежде всего, streamlined. Фигурально и несколько пафосно излагая, я повысил аэродинамическое качество объяснений материала и формулировки заданий. Задания были переработаны и их количество уменьшилось, но отдача от их выполнения увеличилась. Примеры: за три занятия реально изучаются полевой транзистор с примерами его применения и несколько сопутствующих тем [ http://ccfit.nsu.ru/~tregub/eecs/notes12.htm , http://ccfit.nsu.ru/~tregub/eecs/notes13.htm , http://ccfit.nsu.ru/~tregub/eecs/notes14.htm]. Ранее эта тема занимала пять занятий [ http://ccfit.nsu.ru/~tregub/ee_1/eebook_fet.pdf ], и никто из студентов не «проходил» ее далее каскада с общим истоком. В тьюториале, расчитанном на одно занятие в классе и четыре часа самостоятельной работы, удается ознакомить студента со схемотехникой КМОП-защелок (latches) и исследовать явление метастабильности [ http://ccfit.nsu.ru/~tregub/eecs/notesX.htm ]. Этот материал, насколько мне известно, впервые вошел в программу курса электроники в НГУ.

Думаю, что я правильно ответил на вопрос «чему мы учили в курсе унаследованной методики» и ценный учебный материал тех курсов не утрачен в новом курсе. Что касается вопросов о перспективной тематике курса электронной инженерии, практическая учебная работа во время курса отодвинула теоретические размышления, и вопросы пока остались неотвеченными. Некоторые наброски содержатся в материале лекций, которые я читал параллельно с выдачей практического курса [ http://ccfit.nsu.ru/~tregub/eecs/DesignAndAutomation.pdf , http://ccfit.nsu.ru/~tregub/eecs/intro2010.pdf ].